\documentclass[12pt]{article}
\newcommand{\beq}{\begin{eqnarray}}% can be used as {equation} or {eqnarray}
\newcommand{\eeq}{\end{eqnarray}}
\newcommand{\half}{\frac12}
\newcommand{\bi}{\bibitem}

\def\be{\begin{equation}}
\def\ee{\end{equation}}
\def\bea{\begin{eqnarray}}
\def\eea{\end{eqnarray}}

\def\S{{\bf S}}
\title{A Gravity Dual of the Chiral Anomaly}
\author{Igor R. Klebanov$^a$, Peter Ouyang$^a$ and Edward Witten$^b$\\ \\
\small \sl $^a$Department of Physics, Princeton University\\
         \small \sl Princeton, NJ 08544, USA\\
  \small \sl $^b$School of Natural Sciences, Institute for Advanced Studies,\\
\small \sl Princeton, NJ 08540, USA\\ \\
  }

\begin{document}
\setlength{\baselineskip}{16pt}
\begin{titlepage}
\maketitle
\begin{picture}(0,0)(0,0)
\put(325,300){hep-th/0202056}
\put(325,315){PUPT--2022}
\end{picture}
\vspace{-36pt}
\begin{abstract}

We study effects associated with the chiral anomaly for a cascading
$SU(N+M)\times SU(N)$ gauge theory using gauge/gravity duality.  In
the gravity dual the anomaly is a classical feature of the
supergravity solution, and the breaking of the $U(1)$ R-symmetry down to
${\bf Z}_{2M}$ proceeds via the Higgs mechanism.

\end{abstract}
\thispagestyle{empty}
\setcounter{page}{0}
\end{titlepage}

\renewcommand{\baselinestretch}{1.2}  %looks better

\section{Introduction}

Many supersymmetric gauge theories exhibit a classical $U(1)$
R-symmetry which is broken quantum mechanically to some discrete
subgroup.  In traditional quantum field theory, this symmetry breaking
can be understood as an instanton effect.  The purpose of this
note is to explore the analogous effects in the gravity duals to a few
field theories exhibiting this phenomenon.

Our analysis relies on the results of recent encouraging progress in
extending the gauge theory/supergravity correspondence \cite{jthroat,US,EW}
to theories with less than maximal supersymmetry, realized by
configurations of D-branes at singularities.  For example, we might
consider a stack of $N$ coincident D3-branes located at the tip of the
singular Calabi-Yau space known as the conifold \cite{KW}.  This
system is dual to an ${\cal N}=1$ supersymmetric gauge theory with
gauge group $SU(N) \times SU(N)$ coupled to chiral superfields (two
bifundamentals transforming in the (${\bf N}, \overline{{\bf N}}$)
representation, and their conjugates, which transform in the
($\overline{{\bf N}}, {\bf N}$) representation.)  The supergravity
solution has the geometry $AdS_5 \times T^{1,1}$, which manifests
geometrically the conformal invariance of the gauge theory.  The
Einstein manifold $T^{1,1}$ possesses a two-cycle and a three-cycle,
and we can obtain interesting variations of this theory by wrapping
various branes on these cycles \cite{GK,KN,KT,KS}.  In particular,
wrapping D5-branes on this two-cycle changes the gauge group to $SU(N)
\times SU(N+M)$ and breaks the conformal symmetry.

This supergravity solution with wrapped D5-branes exhibits a host of
interesting gauge theory phenomena; for example, the reduction of the
five-form flux as the radial coordinate decreases corresponds to a
reduction in the size of the gauge groups by a duality cascade
\cite{KS}.  An important feature for our purposes is that the UV {\it
metric} exhibits a $U(1)$ symmetry under rotations of a particular
angle $\beta$ in the tranverse space $T^{1,1}$, which is a geometric
realization of the field theoretic R-symmetry.  However, the RR
potentials break the $U(1)$ symmetry. It can be shown that there is an
unbroken ${\bf Z}_{2M}$ subgroup of this $U(1)$ by studying fractional
instanton and domain wall probes in the gravity background
\cite{MN,SL,Gomis}. However, one should not think about this symmetry
breaking as an {\it effect} of these instantons, which do not appear
explicitly anywhere in the gravity dual.  Rather, we argue that the
field theory anomalies are present because the classical supergravity
RR potentials are not invariant under the $U(1)$ symmetry. By
computing the variation of the RR potentials we obtain the relevant
anomaly coefficients, which agree with field theory exactly. Moreover,
the anomalous breaking of the global $U(1)$ symmetry appears as
spontaneous symmetry breaking in supergravity: the bulk vector field
dual to the R-symmetry current of the gauge theory acquires a mass.
We will also check the anomaly coefficients for a related ${\cal N}=2$
orbifold theory, again showing exact agreement.

\section{The Anomaly as Non-Invariance of the UV Supergravity Solution}

Let us recall a few results regarding the supergravity dual of the
cascading $SU(N+M) \times SU(N)$ gauge theory.  The metric is of the
form \cite{KS}
\be \label{specans}
ds^2_{10} =   h^{-1/2}(\tau)   dx^2_{||}
 +  h^{1/2}(\tau) ds_6^2 \ ,
\ee
where $ds_6^2$ is the metric of the deformed conifold.  The UV (large
$\tau$) limit of this metric was found in \cite{KT}:
\be \label{fulsol}
ds^2_{10} =   h^{-1/2}(r)   dx^2_{||}
 +  h^{1/2} (r)  (dr^2 + r^2 ds^2_{T^{1,1}} )\ .
\ee
The metric on $T^{1,1}$, the base of the conifold, is
\begin{equation} \label{co}
ds_{T^{1,1}}^2=
{1\over 9} \bigg(2 d\beta +
\sum_{i=1}^2 \cos \theta_i d\phi_i\bigg)^2+
{1\over 6} \sum_{i=1}^2 \left(
d\theta_i^2 + {\rm sin}^2\theta_i d\phi_i^2
 \right)
\ .
\end{equation}
We define the angle $\beta$ to range from $0$ to $2\pi$; it is related
to the angle $\psi$ used in the previous literature by $\psi = 2
\beta$.  The asymptotic form of the warp factor is
\be
\label{nonharm} h(r) = {27 \pi (\alpha')^2\over 4 r^4} \left [g_s N +
{3\over 2\pi} (g_s M)^2 \ln (r/r_0) + {3\over 8\pi} (g_s M)^2 \right ]
\ .\ee
For sufficiently small $r$, this metric is singular; to study IR physics in the gauge theory, one must use the full solution of \cite{KS}.

The following basis of 1-forms on the compact space is
convenient for calculations on the deformed conifold \cite{MT}:
\bea \label{fbasis}
g^1 = {e^1-e^3\over\sqrt 2}\ ,\qquad
g^2 = {e^2-e^4\over\sqrt 2}\ , \nonumber \\
g^3 = {e^1+e^3\over\sqrt 2}\ ,\qquad
g^4 = {e^2+ e^4\over\sqrt 2}\ , \nonumber \\
g^5 = e^5\ ,
\eea
where
\begin{eqnarray}
e^1\equiv - \sin\theta_1 d\phi_1 \ ,\qquad
e^2\equiv d\theta_1\ , \nonumber \\
e^3\equiv \cos 2\beta \sin\theta_2 d\phi_2-\sin 2\beta d\theta_2\ , \nonumber\\
e^4\equiv \sin 2\beta \sin\theta_2 d\phi_2+\cos 2\beta d\theta_2\ , \nonumber \\
e^5\equiv 2d\beta + \cos\theta_1 d\phi_1+ \cos\theta_2 d\phi_2 \ .
\end{eqnarray}
In terms of this basis, the Einstein metric on $T^{1,1}$ assumes the
form
\be
ds^2_{T^{1,1}}= {1\over 9} (g^5)^2 + {1\over 6}\sum_{i=1}^4 (g^i)^2
\ .
\ee

The three-form fields are turned on in this background:
\be \label{closedf}
F_3 = {M\alpha'\over 2}\omega_3\ ,  \qquad\qquad
B_2 = {3 g_s M \alpha'\over 2}\omega_2 \ln (r/r_0)
\ ,
\ee
\be
H_3 = dB_2 = {3 g_s M \alpha' \over 2r} dr\wedge \omega_2\ ,
\label{hthree}
\ee
where
\beq
\omega_3 &=& g^5 \wedge \omega_2 \, \label{omegathr} \\
%g^5 &=& (2 d\beta + \cos \theta_1 d\phi_1 + \cos \theta_2 d\phi_2) \, \\
\omega_2 &=&
{1\over 2} (\sin\theta_1 d\theta_1 \wedge d\phi_1-
\sin\theta_2 d\theta_2 \wedge d\phi_2 ).
\label{omegato}
\eeq
A key feature of this solution is that the five-form flux is radially
dependent \cite{KT}:
\be F_5={\cal F}_5 +\star
{\cal F}_5 \ , \qquad {\cal F}_5 = \half \pi \alpha'^2 N_{eff}(r)
\omega_2 \wedge \omega_3 \ ,
\label{F5KT}
\ee
with
\be
N_{eff} (r) = N +{3 \over 2\pi} g_s M^2 \ln (r/r_0)
\ ,
\label{Neff}
\ee
and the ten-dimensional Hodge dual is defined by
$\varepsilon_{txyzr5\theta_1 \phi_1 \theta_2 \phi_2} =
\sqrt{-G_{10}}$. In \cite{remarks} it was shown that
\be
\int_{\S^2} \omega_2 = 4 \pi\ , \qquad
\int_{\S^3} \omega_3 = 8 \pi^2 \
\label{intforms}
\ee
The normalization of the R-R 3-form flux is determined by the quantization
condition
\be \label{qcond}
{1\over 4\pi^2 \alpha'}\int_{\S^3} F_3 = M\ .
\ee

With these results in hand, let us see how the chiral anomaly emerges
in supergravity.  The asymptotic UV metric (\ref{fulsol},\ref{co}) has
a $U(1)$ symmetry associated with the rotations of the angular
coordinate $\beta$, which appears as the R-symmetry of the dual gauge
theory.  It is crucial, however, that the background value of the R-R
2-form $C_2$ does not have this continuous symmetry.  Indeed, although
$F_3$ is $U(1)$ symmetric, there is no smooth global expression for
$C_2$. Locally, we may write for large $r$,
\be \label{orig} C_2\rightarrow M\alpha'\beta \omega_2 \ .\ee
This expression is not single-valued as a function of the angular
variable $\beta$, but it is single-valued up to a gauge
transformation, so that $F_3=dC_2$ is single-valued.  In fact, $F_3$
is completely independent of $\beta$.  Because of the explicit $\beta$
dependence, $C_2$ is not $U(1)$-invariant. Under the
transformation $\beta\rightarrow \beta+\epsilon$,
\be \label{shift}
C_2 \rightarrow C_2 + M\alpha' \epsilon \omega_2
\ .\ee
A gauge transformation can shift $C_2/(4\pi^2\alpha')$ by an arbitrary
integer multiple of $\omega_2/(4\pi)$, so $\beta\to\beta+\epsilon$ is
a symmetry precisely if $\epsilon$ is an integer multiple of
$\pi/M$; because $\epsilon$ is anyway only defined mod $2\pi$, a ${\bf Z}_{2M}$ subgroup of the $U(1)$ leaves
fixed the asymptotic values of the fields, and thus corresponds to a
symmetry of the system.
%This ${\bf Z}_{2M}$ is a symmetry since it
%respects the asymptotic values of the fields, but in the solution
%found in \cite{KS}, it is spontaneously broken to ${\bf Z}_2$,
%generated by $(-1)^F$, since the full solution does not have ${\bf
%Z}_{2M}$ symmetry. (In fact, in that solution, ${\bf Z}_{2M}$ is
%broken to ${\bf Z}_2$ by the deformation parameter of the conifold.)
This ${\bf Z}_{2M}$ respects the asymptotic values of the fields, but in the solution
found in \cite{KS}, it is spontaneously broken in the IR to ${\bf Z}_2$,
generated by $(-1)^F$, since the full solution does not have ${\bf
Z}_{2M}$ symmetry. (In that solution, ${\bf Z}_{2M}$ is
broken to ${\bf Z}_2$ by the deformation parameter of the conifold.)
The analogous statement in field theory is that instantons break the
$U(1)$ down to ${\bf Z}_{2M}$, which is then spontaneously broken to
${\bf Z}_2$.  The domain walls interpolating between these $M$
inequivalent vacua are D5-branes wrapped over the 3-sphere at $\tau=0$
in the solution of \cite{KS}, and indeed one such 5-brane produces a
shift of $C_2/(4\pi^2\alpha')$
by exactly $\omega_2/(4\pi)$ \cite{remarks}.

The way that the asymptotic behavior of  $C_2$ transforms under
the $U(1)$ generator is dual to the way that in field theory a
$U(1)_R$ transformation shifts the $\Theta$-angles of the two
gauge groups by opposite amounts.  The $\Theta$-angles are given
by
\be \label{diffsum} \Theta_1 - \Theta_2 ={1\over \pi\alpha'}
\int_{S^2} C_2 \ ,\qquad \Theta_1 + \Theta_2 \sim C \ , \ee
where $C$ is the RR scalar, which vanishes for the case under
consideration. Using the fact that $\int_{\S^2} \omega_2 = 4\pi$, we
find that the small $U(1)$ rotation induces
\be \label{angles}
\Theta_1 =-\Theta_2 = 2 M\epsilon
\ .
\ee

We can compare (\ref{angles}) with our expectations from the field
theory.  The conventionally normalized $\Theta$ terms for the gauge
theory action are
\be
\int d^4 x ({\Theta_1\over 32 \pi^2}
 F^a_{ij} \tilde F^{aij} +  {\Theta_2\over 32 \pi^2}
G^b_{ij} \tilde G^{bij})\ ,
\ee
where $F^a_{ij}$ and $G^b_{ij}$ are the field strengths of $SU(N+M)$
and $SU(N)$ respectively.  If we assume that $\epsilon$ is a function
of the 4 world volume coordinates $x^i$, then the terms linear in
$\epsilon$ in the dual gauge theory are
\be \label{holanom}
\int d^4 x [- \epsilon \partial_i J^i +
{M\epsilon \over 16 \pi^2} (
 F^a_{ij} \tilde F^{aij} -  G^b_{ij} \tilde G^{bij} )
]\ ,
\ee
where $J^i$ is the chiral $R$-current.  The appearance of the second
term is due to the non-invariance of $C_2$ under the $U(1)$ rotation.
Varying with respect to $\epsilon$, we therefore find the equation
\be
 \partial_i J^i = {M \over 16 \pi^2}
(
 F^a_{ij} \tilde F^{aij} -  G^b_{ij} \tilde G^{bij} ) \ .
\label{anomeq}
\ee
This is precisely the anomaly equation for this theory. Indeed, the
effective number of flavors for the $SU(N+M)$ factor is $2N$, and each
one carries R-charge $1/2$. The chiral fermions which are their
superpartners have R-charge $-1/2$ while the gluinos have R-charge
$1$. Therefore, the anomaly coefficient is ${M \over 16 \pi^2}$.  An
equivalent calculation for the $SU(N)$ gauge group with $2(N+M)$
flavors produces the opposite anomaly, in agreement with the
holographic result (\ref{holanom}).

The upshot of the calculation presented above is that the chiral
anomaly of the $SU(N+M)\times SU(N)$ gauge theory is encoded in
the ultraviolet (large $r$) behavior of the dual classical
supergravity solution. No additional fractional D-instanton
effects are needed to explain the anomaly. Thus, as often occurs
in the gauge/gravity duality, a quantum effect on the gauge theory
side turns into a classical effect in supergravity.

%Although the $U(1)$ symmetry is broken, there is a ${\bf Z}_{2M}$
%discrete subgroup of the $U(1)$ which remains a symmetry. From the
%field theory point of view, this is due to the periodicity under
%$\Theta_i \rightarrow \Theta_i + 2\pi k_i$.  Thus, in view of
%(\ref{angles}), $\beta \rightarrow \beta+ \pi k/M$ corresponds to
%a discrete symmetry transformation. Since the period of $\beta$ is
%$2\pi$, we find the unbroken ${\bf Z}_{2M}$ as expected.  We
%obtained the same result above in string theory using the fact
%that  ${1\over 2\pi\alpha'}\int_{\S^2} C_2$ is a periodic variable
%with period $2\pi$. In the IR completion of the solution found in
%\cite{KS}, one sees that a D5-brane wrapped over the 3-sphere at
%$\tau=0$ produces a shift of ${1\over 2\pi\alpha'}\int_{\S^2}
%C_2$ by exactly one period \cite{remarks}.  The solution of
%\cite{KS} exhibits the breaking of ${\bf Z}_{2M}$ to ${\bf Z}_2$
%in the IR. Thus, there are $M$ inequivalent vacua, and the
%wrapped D5-brane is a domain wall separating two adjacent vacua.

\section{The Anomaly as Spontaneous Symmetry Breaking in $AdS_5$}

Let us look for a deeper understanding of the anomaly from the
dual gravity point of view.  On the gauge theory side, the
R-symmetry is global, but in the gravity dual it as usual becomes
a gauge symmetry, which must not be anomalous, or the theory
would not make sense at all.  Rather, we will find that the gauge
symmetry is spontaneously broken: the 5-d vector field dual
to the R-current of the gauge theory `eats' the scalar dual to
the difference of the theta angles and acquires a mass.
\footnote{ The connection between
anomalies in a D-brane field theory and spontaneous symmetry
breaking in string theory was previously noted in \cite{aks} (and
probably elsewhere in the literature).}
%but we are not aware of
%work on its realization in the AdS/CFT correspondence.
A closely related mechanism was observed in studies of RG flows
from the dual gravity point of view \cite{BDFP,brand,BFS}. There R-current
conservation was violated not through anomalies but by turning on
relevant perturbations or expectation values for fields. In these
cases it was shown \cite{BDFP,brand,BFS}
that the 5-d vector field dual to the R-current
acquires a mass through the Higgs mechanism. We will show that
symmetry breaking through anomalies can also have the bulk Higgs mechanism
as its dual.

In the absence of fractional branes there are no background three-form
fluxes, so the $U(1)$ R-symmetry is a true symmetry of the field
theory.  Because the R-symmetry is realized geometrically by
invariance under a rigid shift of the angle $\beta$, it becomes a
local symmetry in the full gravity theory, and the associated gauge
fields $A = A_{\mu} dx^{\mu}$ appear as fluctuations of the
ten-dimensional metric and RR four-form potential
\cite{krvn,ceresole}.  The natural metric ansatz is of the familiar
Kaluza-Klein form:
\beq
  ds^2 = h(r)^{-1/2} \left( dx_n dx^n \right) +
  h(r)^{1/2} r^2 \left[\frac{dr^2}{r^2} +
\frac{1}{9} \left(g^5 - 2A \right)^2+
          \frac{1}{6} \sum_{r=1}^4 \left( g^{r}\right)^2 \right],
\label{kkmetric}
\eeq
where $h(r) = L^4/r^4$, and $L^4 = \frac{27}{16}(4\pi \alpha'^2 g_s
N)$.  It is convenient to define the one-form $\chi = g^5 -
2A$, which is invariant under the combined gauge transformations
\beq
\beta \rightarrow \beta + \lambda,
\qquad A \rightarrow A + d\lambda.
\label{gaugetrans}
\eeq
The equations of motion for the field $A_{\mu}$ appear as the $\chi\mu$
components of Einstein's equations,
\beq
   R_{MN} = \frac{g_s^2}{4 \cdot 4!} \tilde{F}_{MPQRS} \tilde{F}_N^{~PQRS}.
\eeq

The five-form flux will also fluctuate when we activate the
Kaluza-Klein gauge field; indeed, the unperturbed $\tilde{F}_5$ of
(\ref{F5KT}) is not self-dual with respect to the gauged metric
(\ref{kkmetric}).  An appropriate ansatz to linear order in $A$ is
\beq \label{fiveform}
   \tilde{F}_5= dC_4 &=& \frac{1}{g_s} d^4 x \wedge dh^{-1} +
   \frac{\pi \alpha'^2 N}{4} \Bigg[\chi \wedge
   g^1 \wedge g^2 \wedge g^3 \wedge g^4 \nonumber \\
  & & \qquad \qquad - dA \wedge g^5 \wedge dg^5 +
  \frac{3}{L} \star_5 dA \wedge dg^5 \Bigg].
\eeq
The five-dimensional Hodge dual $\star_5$ is defined with respect to
the AdS$_5$ metric $ds_5^{~2}=h^{-1/2} dx_n dx^n + h^{1/2} dr^2$.  It
is straightforward to show that the supergravity field equation $
d\tilde F_5 =0$ implies that the field $A$ satisfies the equation of
motion for a massless vector field in AdS$_5$ space:
\beq d \star_5 dA =0 \ .
\label{vecteq}
\eeq
Using the identity $dg^5 \wedge dg^5 = -2 g^1 \wedge g^2 \wedge g^3
\wedge g^4$, we can check that the expression for $C_4$ is\footnote{
This expression was independently derived by D. Berenstein.}
\beq
C_4 &=& {1\over g_s} h^{-1} d^4 x + \frac{\pi \alpha'^2 N}{2}
\Bigg[ \beta g^1 \wedge g^2 \wedge g^3 \wedge g^4 - {1\over 2}
A\wedge d g^5 \wedge g^5 \nonumber \\
& & \qquad \qquad \qquad -\frac{3}{2r} h^{-1/4} \star_5 dA \wedge g^5 \Bigg].
\label{cfourads}
\eeq
Another way to see that $A$ is a massless vector in $AdS_5$ is to
consider the Ricci scalar for the metric (\ref{kkmetric})
\beq
  R = R(A=0) - \frac{h^{1/2}r^2}{9} F_{\mu \nu} F^{\mu \nu}
\label{riccikk}
\eeq
so that on reduction from ten dimensions the five-dimensional
supergravity action will contain the action for a massless vector
field.

The story changes when we add wrapped D5-branes.  As described in
Section 2, the 5-branes introduce $M$ units of RR flux through the
three-cycle of $T^{1,1}$.
%In what follows, we work in the limit $(g_s
%M)^2 \ll g_s N$, so that it is consistent to expand the metric and
%five-form to quadratic order in $g_s M$ and expand the three-forms and
%one-forms to linear order in $g_s M$.  At this order we may
%consistently take the dilaton and axion to vanish.
Now,
%with the
%addition of 5-branes
the new wrinkle is that the RR three-form flux
of (\ref{closedf}) is not gauge-invariant with respect to shifts of
$\beta$ (\ref{gaugetrans}).
To restore the gauge invariance, we introduce a new field $\theta\sim
\int_{S^2} C_2$:
\beq
F_3 = dC_2 = \frac{M\alpha'}{2} \left( g^5 +
              2 \partial_{\mu} \theta dx^{\mu} \right) \wedge \omega_2
\eeq
so that $F_3$ is invariant under the gauge transformation $\beta
\rightarrow \beta + \lambda$, $\theta \rightarrow \theta -\lambda$.
Let us also define $W_{\mu} = A_{\mu} + \partial_{\mu} \theta$.  In
terms of the gauge invariant forms $\chi$ and $W=W_{\mu}dx^{\mu}$,
\beq
F_3 = \frac{M\alpha'}{2} (\chi + 2W) \wedge \omega_2.
\label{f3shift}
\eeq

{}From (\ref{f3shift}) we can immediately see how the anomaly will
appear in the gravity dual.  Assuming that the NS-NS three form is
still given by (\ref{hthree}), we find that up to terms of order
$g_sM^2/N$ the three-form equation implies
\beq
d\star_5 W = 0 \, \Rightarrow \, \frac{L^2}{r^2} \partial_i W^i +
\frac{1}{r^5} \partial_r r^5 W_r = 0
\label{dstarw}
\eeq
which is just what one would expect for a massive vector field in five
dimensions.  To a four dimensional observer, however, a massive vector
field would satisfy $\partial_i W^i=0$.  Thus in the field theory one
cannot interpret the $U(1)$ symmetry breaking as being spontaneous,
and the additional $W_r$ term in (\ref{dstarw}) appears in four
dimensions to be an anomaly.

Another way to see that the vector field becomes massive is to compute
its equation of motion.  To do this calculation precisely, we should
derive the $\chi \mu$ components of Einstein's equations, and also
find the appropriate expressions for the five-form and metric up to
quadratic order in $g_s M$ and linear order in fluctuations.  This
approach is somewhat nontrivial.
%and we present it in an Appendix for
%the interested reader.
A more heuristic approach is
to consider the type IIB supergravity action to quadratic order in
$W$, ignoring the 5-form field strength contributions:
\beq
   S & = & -\frac{1}{2\kappa_{10}^{~2}} \int d^{10}x \sqrt{-G_{10}}
\left[ R_{10} -\frac{g_s^{~2}}{12}|F_3|^2 \right] + \ldots \\
    &\sim& -\frac{1}{2\kappa_{10}^{~2}} \int d^{10}x \sqrt{-G_{10}}
    \left[ - \frac{h^{1/2}r^2}{9} F_{\mu \nu} F^{\mu \nu} -
    \left(\frac{g_s M\alpha'}{2} \right)^2
     \frac{36}{hr^4} W_{\mu} W^{\mu} \right] + \ldots
\label{action}
\eeq
This is clearly the action for a massive vector
field, which has as its equation of motion
\beq
\partial_{\mu} (hr^7 F^{\mu \nu}) = \tilde m^2 hr^7 W^{\nu}
\label{weom}
\eeq
which in differential form notation is $d(h^{7/4}r^7\star_5 dW) = -
\tilde m^2
h^{7/4}r^7 \star_5 W$.  Here
the mass-squared is given by
\beq
\tilde m^2 &=& \left(g_s M\alpha' \right)^2 \frac{81}{2h^{3/2} r^6}.
\label{mass}
\eeq
%
%In the limit $(g_s M)^2 \ll g_s N$ which we are using, this becomes
%\beq
% \tilde m^2 = \frac{4}{\alpha' 3^{1/2}
%\pi^{3/2} } \frac{(g_s M)^2}{(g_s N)^{3/2}}.
%\label{newmass}
%\eeq
This result, however, ignores the subtlety of the type IIB action in
presence of the self-dual 5-form field.
A more precise calculation \cite{krasnitz}, which takes the mixing into account,
gives instead the following equation for the transverse vector modes:
\beq \label{mk}
\left ({1 \over hr^7}\partial_r hr^7 \partial_r +
h\partial_{i}\partial_{i}-{(9 M\alpha')^4 \over 64 h^2 r^{10} }\right )W_{i}=0,
\eeq
This shows that the 10-d mass actually appears at a higher order in
perturbation theory compared to the result (\ref{mass})
that ignores the mixing
with the 5-form.\footnote{We are grateful to M. Krasnitz for
correcting an error in a previous version of this paper.}

Let us compare this result to earlier work.
In \cite{BDFP,brand,BFS} it was shown that the 5-d vector
field associated with a $U(1)_R$ symmetry acquires a mass in the
presence of a symmetry-breaking relevant perturbation,
and that this mass is related in a simple way to the warp factor of the
geometry.\footnote{We are grateful to O.~DeWolfe and K.~Skenderis for pointing
out the relevance of this work to the present calculation.}
It is conventional to write the 5-d gauged supergravity
metric
in the form
\beq
\tilde{G}_{\mu \nu} dx^{\mu}dx^{\nu} = e^{2T(q)}\eta_{ij} dx^i dx^j +dq^2.
\label{dwmetric}
\eeq
The result of \cite{BDFP} is that $m^2 = -2T''$.  To relate
the 5-d metric (\ref{dwmetric}) to the 10-d metric (\ref{specans})
we must normalize the 5-d metric so that the graviton
has a canonical kinetic term.  Doing this carefully we find
\beq
\tilde{G}_{\mu \nu}dx^{\mu}dx^{\nu} =
\left( hr^4/L^4 \right)^{5/6} (h^{-1/2}\eta_{ij} dx^i dx^j+h^{1/2}dr^2).
\label{metric5d}
\eeq
The factor $\left( hr^4/L^4 \right)^{5/6}$ arises due to the
radial dependence of the size of $T^{1,1}$ through the
usual Kaluza--Klein reduction.
The radial variables $q$ and $r$ are related
at leading order in $g_s M^2/N$, by
\beq
\log(r) \sim \frac{q}{L} - \frac{g_sM^2}{2\pi N}\left(\frac{q}{L}\right)^2.
\eeq
We can also show that $-2T = -2\log(r) +$(terms which do not affect
the mass to leading order in $g_s M^2/N$),
so now computing the mass-squared
by the prescription of \cite{BDFP} we
obtain
\beq
  m^2 = \frac{4}{\alpha' (3\pi)^{3/2} } \frac{(g_s M)^2}{(g_s N)^{3/2}}.
\label{rightmass}
\eeq
where this mass applies to a vector field $V$ with a canonical
kinetic term for the metric (\ref{metric5d}).
For these calculations it is convenient to work with the transverse
4-d vector modes $V_i$ and to decouple the longitudinal modes such as $V_r$.
The equation of motion of $V$ is
\beq
(e^{-2T} \frac{\partial}{\partial q}e^{2T} \frac{\partial}{\partial q} +
e^{-2T}\partial_i \partial_i - m^2)V_i = 0.
\eeq
In fact, this equation follows from (\ref{mk}) after a rescaling
\cite{krasnitz}
\beq
V_i = (hr^4/L^4)^{2/3} W_i.
\eeq
The nonvanishing vector mass is consistent with gauge invariance because
the massless vector field $A$ has eaten the scalar field $\theta$,
spontaneously breaking the gauge symmetry, as advertised.  It is
interesting that the anomaly appears as a bulk effect in
AdS space, in contrast to some earlier examples
\cite{EW,henningson} where anomalies arose from boundary terms.

The appearance of a mass implies that the R-current operator should
acquire an anomalous dimension.
{}From (\ref{rightmass}) it follows that
\be (m L)^2 = \frac{2 (g_s M)^2}{\pi (g_s N)}\ .
\ee
Using the AdS/CFT correspondence (perhaps naively, as the KT solution is not asymptotically AdS)
we find that the dimension of the
current $J^\mu$ dual to the vector field $W^\mu$ is
\be
\Delta = 2 + \sqrt{1 + (mL)^2} \ .
\ee
Therefore, the anomalous dimension of the current is
\be
\Delta - 3 \approx (mL)^2/2 =\frac{(g_s M)^2}{\pi (g_s N)}\ .
\label{anomdim}
\ee
We can obtain a rough understanding of this result by considering the
relevant weak coupling calculation in the gauge theory.  The leading
correction to the current-current two-point function comes from the
three-loop Feynman diagram composed of two triangle diagrams glued
together, and the resulting anomalous dimension $\gamma_J$ is
quadratic in $M$ and $N$.  $\gamma_J$ must vanish when $M=0$, and it
must be invariant under the map $M \rightarrow -M$, $N \rightarrow
N+M$, which simply interchanges the two gauge groups.  Thus, the
lowest order piece of the anomalous dimension will be of order $(g_s
M)^2$.  Our supergravity calculation predicts that this anomalous
dimension is corrected at large $g_s N$ by an extra factor of $1/(g_s
N)$.  Of course, it would be interesting to understand this result
better from the gauge theory point of view.

\section{The ${\cal N}=2$ Supersymmetric ${\bf Z}_2$ Orbifold}

Encouraged by the agreement of field theory and supergravity on
the conifold, let us examine another example to see how the same
physical ideas apply in a different system.  In this section we
will study the ${\cal N}=2$ version of the conifold theory; it
has gauge group $SU(N+M) \times SU(N)$ and is dual to a
supergravity solution on an orbifold $S^5/{\bf Z}_2$
\cite{GK,KN,joe,bert}. (After the completion of this work, we learned of
a very similar analysis of this orbifold system which appeared earlier in
\cite{bertd7}.)  The supergravity solution may be
constructed as follows.  We start with the space ${\bf R}^{1,5}
\times {\bf R}^4/{\bf Z}_2$ where the orbifold is given by the
identification $x_{6,7,8,9} \sim -x_{6,7,8,9}$.  Then we add $N$
coincident D3-branes, which we choose to be tangent to the 0123
directions; the resulting space has the geometry $AdS_5 \times
S^5/{\bf Z}_2$.  To add fractional branes, we may take $M$
D5-branes and wrap them on the vanishing two-cycle of the
orbifold ${\bf R}^4/{\bf Z}_2$.  These fractional branes are
``pinned'' to the orbifold fixed plane.

It is possible to identify the corresponding gauge theory by standard
orbifold techniques \cite{douglasmoore}.  The field content is in fact
almost identical to that of the conifold theory, but there is an
additional pair of adjoint chiral multiplets corresponding to the
motion of D-branes along the orbifold fixed plane.  These extra
multiplets combine with the vector multiplet in the ${\cal N} =1$
theory to form an ${\cal N}=2$ vector multiplet.  It is convenient to
define $(x_4 + ix_5)/(2\pi \alpha') \equiv \Phi =|\Phi| e^{i\beta}$.
Rotations of the phase of $\Phi$ are dual to the $U(1)$ R-symmetry in
the gauge theory.

To compute the anomaly for the $SU(N+M)$ gauge factor, notice that
there are now $2N+(N+M)=3N+M$ effective flavors, whose
fermionic components have R-charge
$-1/3$.  Combining this with the contribution from the gluinos, we find
that the anomaly coefficient is $\frac{2M/3}{16\pi^2}$.  We would like
to compare this with a computation from supergravity.  Equations
(\ref{orig}),(\ref{shift}) are also satisfied for the orbifold.  To
identify properly the relation between $\beta$ and the R-symmetry,
note that the field $\Phi$ has R-charge 2/3; thus a shift of $\beta
\rightarrow \beta +\epsilon$ actually shifts the $U(1)_R$ by $\frac32
\epsilon$.  This will change the first term in (\ref{holanom}) by a
factor of $\frac32$ and give an anomaly coefficient
$\frac{2M/3}{16\pi^2}$ in agreement with the gauge theory expectation.

A very interesting generalization of this theory was studied
by Gra\~{n}a and Polchinski \cite{Grana}, and also by Bertolini et al
\cite{bertd7}, who added D7-branes wrapped
on the 01236789 directions; an analogous solution with D7-branes on
the conifold is not currently known.  The extra D7-branes allow
excitations of 3-7 strings and, depending on how the 7-branes are
wrapped, will add $N_{7+}$ flavors coupled to the $SU(N+M)$ gauge
group and $N_{7-}$ flavors coupled to the $SU(N)$ gauge group.  The
total number of 7-branes is $N_{7+} + N_{7-}$.
The fermionic components of
these flavors also have R-charge $-1/3$.  For a small rotation
$\beta \rightarrow \beta + \epsilon$, the corresponding $\Theta$ terms
are
\beq
\Theta_1 & = & 2\epsilon (M - \half N_{7+}) \label{t1flavors}\\
\Theta_2 & = & 2\epsilon (-M - \half N_{7-})\label{t2flavors}.
\eeq
and the associated anomaly coefficient is $\frac{1}{16 \pi^2} (2M/3 -
N_{7+}/6 +N_{7-}/6)$.

We can reproduce the same result for the anomaly by a supergravity
calculation, using the results of \cite{joe} and \cite{Grana, bertd7}.  It is
helpful to think about the D3-branes on the orbifold fixed plane as a
combination of a wrapped D5-brane and anti-D5-brane, each of which
carries half a unit of D3-brane charge.  By considering the
Chern-Simons term in the action for a probe 5-brane
\beq
\pm \mu_5 \int (2\pi \alpha'){\cal F}^{\pm}_2 \wedge C_4 =
     \pm \frac{\mu_3}{2 \pi} \int{\cal F}^{\pm}_2 \wedge C_4
      = + \half \mu_3 \int C_4,
\eeq
we see that the field strength on the 5-brane worldvolume will satisfy
\beq
{\cal F}^{\pm}_2 = \pm \frac14 \omega_2
\eeq
where the upper sign refers to a D5 and the lower sign to an anti-D5.
Now let us add the D7-branes. The supergravity solution has RR scalar
and two-form potentials given by \cite{Grana,bertd7}
\beq
C_0 & = & -\frac{\beta}{2 \pi} (N_{7+} + N_{7-}) \\
C_2 & = & \alpha' \beta \omega_2 (M - \frac{N_{7+} - N_{7-}}4).
\label{d7fields}
\eeq
We have chosen the signs in (\ref{d7fields}) to differ from those of \cite{bertd7,Grana,moreanom} by a conventional overall minus sign.
To find the $\Theta$ terms for the dual gauge theory, we just need to
look at the Chern-Simons terms in the actions for a probe D5-brane and
anti-D5-brane.  For a D5-brane (whose excitations are in the $SU(N+M)$
gauge group) we find that
\beq
\frac1{2\pi \alpha'}\int
\left( C_2 + 2\pi \alpha' C_0 {\cal F}^+_2 \right) = 2\beta (M-\half N_{7+}).
\label{d7coeff}
\eeq
Comparison with (\ref{diffsum}) and (\ref{angles}) shows that gravity
reproduces the field theory expectation for
$\Theta_1$ given in equation (\ref{t1flavors}).
The analogous computation for an
anti-D5-brane will reproduce (\ref{t2flavors}).  Thus the anomaly as
computed from supergravity agrees exactly with the field theory
calculation.

%%%%%%%%%%%%%%%%%%%%%%%%%%%%%%%%%%%%%%%%%%%%%%%
\section*{Acknowledgements}
%%%%%%%%%%%%%%%%%%%%%%%%%%%%%%%%%%%%%%%%%%%%%%%%%%%%
We are grateful to D. Berenstein, O. DeWolfe,
C. Herzog, M. Krasnitz, K. Skenderis and D.  Vaman
for useful discussions.  I.R.K. also thanks the High Energy Theory
Group at Rutgers University for hospitality while part of this work
was carried out.  This research was supported in part by the NSF Grants
PHY-9802484 and PHY-0070928, and an NSF Graduate Research Fellowship.


\begin{thebibliography}{99}

\bibitem{jthroat}
J.~Maldacena, ``The Large N limit of superconformal field theories and
supergravity,'' {\em Adv. Theor. Math. Phys.} {\bf 2} (1998) 231,
{{\tt hep-th/9711200}}.
%%CITATION = HEP-TH 9711200;%%


\bibitem{US}
S.~S.~Gubser, I.~R.~Klebanov, and A.~M.~Polyakov, ``Gauge theory correlators
  from noncritical string theory,''
{\em Phys. Lett.} {\bf B428} (1998) 105,
 {{\tt hep-th/9802109}}.
%%CITATION = HEP-TH 9802109;%%

\bibitem{EW}
E.~Witten, ``Anti-de Sitter space and holography,''
{\em Adv. Theor. Math. Phys.} {\bf 2} (1998) 253,
{{\tt hep-th/9802150}}.
%%CITATION = HEP-TH 9802150;%%

\bibitem{KW}
I.~R.~Klebanov and E.~Witten,
``Superconformal field theory on threebranes at a Calabi-Yau  singularity,''
{\em Nucl.\ Phys.\ B} {\bf 536}, (1998) 199,
{\tt hep-th/9807080}.
%%CITATION = HEP-TH 9807080;%%


\bi{GK}
S.~S.~Gubser and I.~R.~Klebanov,
``Baryons and Domain Walls in an N = 1 Superconformal Gauge Theory,''
{\em Phys. Rev.} {\bf D58} (1998) 125025,
{{\tt hep-th/9808075}}.
%%CITATION = HEP-TH 9808075;%%

\bi{KN}
I.~R.~Klebanov and N.~Nekrasov,
``Gravity Duals of Fractional Branes and Logarithmic RG Flow,''
{\em Nucl. Phys.} {\bf B574} (2000) 263,
{\tt hep-th/9911096}.
%%CITATION = HEP-TH 9911096;%%

\bi{KT}
I.~R.~Klebanov and A.~Tseytlin,
``Gravity Duals of Supersymmetric $SU(N)\times SU(N+M)$ Gauge Theories,''
{\em Nucl. Phys.} {\bf B574} (2000) 123,
{\tt hep-th/0002159}.
%%CITATION = HEP-TH 0002159;%%

\bi{KS}
I.~R.~Klebanov and M.~Strassler, ``Supergravity and a Confining Gauge
Theory: Duality Cascades and $\chi$SB--Resolution of Naked
Singularities,'' {\em JHEP} {\bf 0008} (2000) 052, {\tt
hep-th/0007191}.
%%CITATION = HEP-TH 0007191;%%

\bi{MN}
J.~Maldacena and C.~Nunez, ``Towards the large $N$ limit of pure
${\cal N}=1$ super Yang Mills,''
{\it Phys. Rev. Lett.} {\bf 86} (2001) 588,
{\tt hep-th/0008001}.
%%CITATION = HEP-TH 0008001;%%

\bi{SL}
A.~Loewy and J.~Sonnenschein,
``On the holographic duals of ${\cal N}=1$ gauge dynamics,''
{\it JHEP} {\bf 0108} (2001) 007,
{\tt hep-th/0103163}.
%%CITATION = HEP-TH 0103163;%%

\bi{Gomis}
Jaume Gomis, ``On SUSY Breaking and ChiSB from String Duals,''
{\tt hep-th/0111060}.
%%CITATION = HEP-TH 0111060;%%

\bi{MT}
R.~Minasian and D.~Tsimpis,
``On the Geometry of Non-trivially Embedded Branes,''
{\em Nucl. Phys.} {\bf B572}, (2000), 499.
{\tt hep-th/9911042}.
%%CITATION = HEP-TH 9911042;%%

\bi{remarks}
C.~P.~Herzog, I.~R.~Klebanov and P.~Ouyang,
``Remarks on the warped deformed conifold,''
{\tt hep-th/0108101}.
%%CITATION = HEP-TH 0108101;%%

\bi{aks}
O.~Aharony, S.~Kachru and E.~Silverstein,
``New N=1 Superconformal Field Theories in Four Dimensions from D-brane
Probes,'' {\em Nucl. Phys.} {\bf B488}, (1997) 159, {{\tt hep-th/9610205}}.
%%CITATION = HEP-TH 9610205;%%

\bi{BDFP}
M.~Bianchi, O.~DeWolfe, D.~Z.~Freedman and K.~Pilch,
``Anatomy of two holographic renormalization group flows,''
{\it JHEP} {\bf 0101}, (2001) 021,
{\tt hep-th/0009156}.
%%CITATION = HEP-TH 0009156;%%

\bi{brand}
A.~Brandhuber and K.~Sfetsos,
``Current correlators in the Coulomb branch of N = 4 SYM,''
{\it JHEP} {\bf 0012}, (2000) 014,
{\tt hep-th/0010048}.
%%CITATION = HEP-TH 0010048;%%

\bi{BFS} M. Bianchi, D.Z. Freedman and K. Skenderis,
``Holographic Renormalization,''
{\tt hep-th/0112119}.
%%CITATION = HEP-TH 0112119;%%

\bi{krasnitz}
M.~Krasnitz, paper to appear.

\bi{krvn}
H.~J.~Kim, L.~J.~Romans and P.~van Nieuwenhuizen,
``The Mass Spectrum of Chiral N=2 D=10 Supergravity on S**5,''
{\em Phys. Rev.} {\bf D32} (1985) 389.
%%CITATION = PHRVA,D32,389;%%

\bibitem{ceresole}
A.~Ceresole, G.~Dall'Agata, R.~D'Auria and S.~Ferrara,
``Spectrum of type IIB supergravity on AdS(5) x T(11):
Predictions on N=1 SCFT's,'' {\em Phys. Rev.} {\bf D61} (2000) 066001,
{\tt hep-th/9905226};
%%CITATION = HEP-TH 9905226;%%
A.~Ceresole, G.~Dall'Agata and R.~D'Auria,
``KK spectroscopy of type IIB supergravity on AdS(5) x T(11),'' JHEP {\bf 9911}, 009 (1999) {\tt hep-th/9907216}.
%%CITATION = HEP-TH 9907216;%%

\bi{henningson}
M.~Henningson and K.~Skenderis,
``The holographic Weyl anomaly,''
JHEP {\bf 9807}, 023 (1998)
{\tt hep-th/9806087}.
%%CITATION = HEP-TH 9806087;%%

\bibitem{joe}
J.~Polchinski,
``N = 2 gauge-gravity duals,''
Int.\ J.\ Mod.\ Phys.\ A {\bf 16}, (2001) 707
{\tt hep-th/0011193}.
%%CITATION = HEP-TH 0011193;%%

\bi{bert}
M.~Bertolini, P.~Di Vecchia, M.~Frau, A.~Lerda, R.~Marotta and I.~Pesando,
``Fractional D-branes and their gauge duals,''
JHEP {\bf 0102}, 014 (2001)
{\tt hep-th/0011077}.
%%CITATION = HEP-TH 0011077;%%


\bibitem{bertd7}
M.~Bertolini, P.~Di Vecchia, M.~Frau, A.~Lerda and R.~Marotta,
``N = 2 gauge theories on systems of fractional D3/D7 branes,''
{\it Nucl.\ Phys.}  {\bf B621}, (2002) 157
{\tt hep-th/0107057}.
%%CITATION = HEP-TH 0107057;%%



\bibitem{douglasmoore}
M.~R.~Douglas and G.~W.~Moore,
``D-branes, Quivers, and ALE Instantons,''
{\tt hep-th/9603167}.
%%CITATION = HEP-TH 9603167;%%

\bibitem{Grana}
M.~Gra\~{n}a and J.~Polchinski,
``Gauge / gravity duals with holomorphic dilaton,''
{\tt hep-th/0106014}.
%%CITATION = HEP-TH 0106014;%%

\bibitem{moreanom}
M.~Bertolini, P.~Di Vecchia, M.~Frau, A.~Lerda and R.~Marotta,
``More Anomalies from Fractional Branes,''
{\tt hep-th/0202195}.
%%CITATION = HEP-TH 0202195;%%

\end{thebibliography}
\end{document}